\documentclass[letterpaper]{article}
\usepackage{aaai18}
\usepackage{times}
\usepackage{helvet}
\usepackage{courier}
\usepackage{url}
\usepackage{graphicx}
\usepackage{subfig}

\usepackage{booktabs}
\usepackage{multirow}

\usepackage{microtype}
\usepackage{xcolor}

\title{Self-Representation on Twitter Using Emoji Skin Color Modifiers}

\author{Alexander Robertson, Walid Magdy, Sharon Goldwater\\
Institute for Language, Computation and Cognition\\
School of Informatics, University of Edinburgh\\
alexander.robertson@ed.ac.uk, \{wmagdy, sgwater\}@inf.ed.ac.uk
}

\begin{document}
\maketitle
\begin{abstract}
Since 2015, it has been possible to modify certain emoji with a skin tone. The five different skin tones were introduced with the aim of representing more human diversity, but some commentators feared they might be used as a way to negatively represent other users/groups. This paper presents a quantitative analysis of the use of skin tone modifiers on emoji on Twitter, showing that users with darker-skinned profile photos employ them more often than users with lighter-skinned profile photos, and the vast majority of skin tone usage matches the color of a user's profile photo---i.e., tones represent the self, rather than the other. In the few cases where users do use opposite-toned emoji, we find no evidence of negative racial sentiment. Thus, the introduction of skin tones seems to have met the goal of better representing human diversity.
\end{abstract}

\section{Introduction}

Emoji---icons used to represent emotions, ideas, or objects---became a formally recognized component of the Unicode Standard in 2010. At that time, all emoji depicting humans were rendered with the same yellow skin tone. However, in 2015, special modifier codes were introduced to allow users to change the skin tone of certain emoji. Unicode Technical Standard \#51 \cite{tsr51}, justified this on the grounds that ``people all over the world want to have emoji that reflect more human diversity.''  Nevertheless, the change was not uncontroversial.  Although skin tone modifiers can be used to make emoji more personal to reflect the identity of the user, in principle they can also be used to represent others. This possibility sparked fears in the mainstream media that users might engage in ``digital blackface'', negatively using dark emoji for entertainment purposes \cite{bbcblackface,techcrunchblackface}---if people would even use them at all \cite{guardiandiverse}. But despite this media and popular attention, and considerable academic investigation of emoji in general, we are not aware of any quantitative studies of how people actually use skin tone modifiers.


In this paper, we present an analysis of how skin tone modifiers are used in emoji on Twitter. In a sample of 0.6 billion tweets, we find that 42\% of the 13 million tone-modifiable emoji (TME) tweeted in 2017 included skin tone modifiers (henceforth, we refer to these modified TME as TME+, in contrast to unmodified TME-) and the majority of these TME+ were light colors. By annotating the profile photos of 4,099 users and analyzing their tweet histories, we show that the overall prevalence of lighter skin tone modifiers is likely due to the prevalence of lighter-skinned users. Indeed, users with darker-skinned profile photos are more likely to use TME, and more likely to modify them. Moreover, the vast majority of TME+ are similar in tone to the user's profile photo, suggesting that TME+ are used overwhelmingly for self-representation. Finally, we analyze the sentiment of the small proportion of tweets containing opposite-toned emoji, and find that positive tweets outweigh negative. Overall, our findings suggest that the Unicode Consortium has successfully met a real demand for diversity within emoji, with few negative consequences.




\section{Emoji and Skin Tones}

\begin{figure}
\center
\includegraphics[width=\columnwidth]{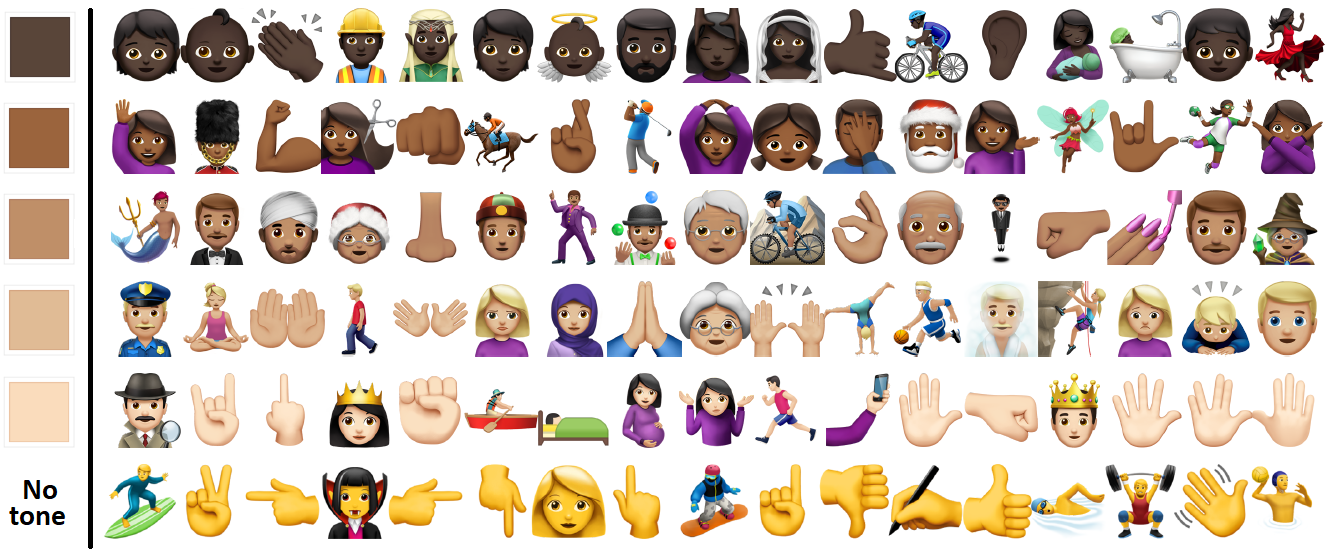}
\caption{The current 102 tone-modifiable emoji. The first five rows are modified by a skin tone (shown in the left-most column), and the final row contains unmodified emoji.}
\label{102emoji}
\end{figure}

\begin{figure*}
\centering
\includegraphics[width=\textwidth]{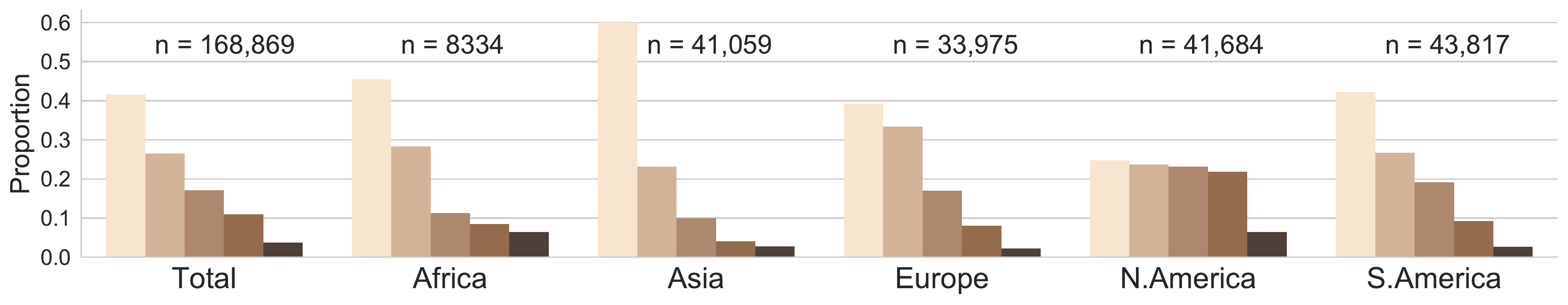}
\caption{Proportion of TME+ using each skin tone, by geographic region.}
\label{regional_usage}
\end{figure*}

Recent years have seen significant academic interest in emoji. Studies show that emotional interpretation of some emoji is subject to ambiguity \cite{miller2017}, that addition of emoji to messages (particularly neutral ones) can convey positive/negative tone and sentiment \cite{intent2017}, and that popularity differences between emoji may be due to their relation to popular words \cite{ai2017untangling}. Resources commonly available for traditional language such as sense inventories have been compiled \cite{emojinet2017} and emoji usage has been linked to personality traits \cite{MARENGO201774}. But despite three years since their introduction, TME have been unstudied until now.

Not all emoji can have a skin tone. The traditional emoji faces (e.g. \includegraphics[scale=0.08]{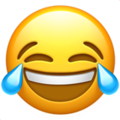}) cannot be modified. The current Unicode Standard defines 102 TME, some of which can be combined to create compound emoji (e.g. \includegraphics[width=0.3cm]{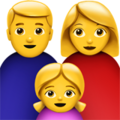}) which can also be modified for skin color. Five skin tones, based on the six tones of the Fitzpatrick phototyping scale \cite{fitzpatrick}, are available. We will refer to the five-scale tones as T1 (the lightest) to T5 (the darkest). Figure \ref{102emoji} illustrates the 102 available TME, showing how the tones affect them.


Emoji are rendered differently across platforms, which can cause ambiguity in semantic interpretation---e.g., writers/readers may perceive the expressed action or emotion differently \cite{miller2017}. Since we are only concerned with skin color, we do not expect these ambiguities to be an issue in our study. However, there is slight variation in the rendering of specific skin tones across platforms. We therefore cannot be sure of the precise tonal values observed by a user, although all platforms have exactly 5-scale tones, where T1 is lighter than T2, T2 lighter than T3, and so forth.

\section{Skin Tone Usage, Overall and By Region}

For our initial analysis we used the Twitter API (1\% sample) to create Dataset A, from 1.04 billion tweets made between February and December 2017. This dataset contains 595m original (non-retweet) tweets made by 769m unique users. Of these, 83.3m (14\%) contain at least one emoji and 13.1m (2.2\%) contain at least one TME. 5.5m (42\%) of TME were TME+. Lighter tones dominate: T1 and T2 account for 68\% of all TME+.

To gain a global view of TME usage by location, we used the Google Maps API to validate the location field of the authors of the 5.5m TME+ tweets \footnote{Though this field is not necessarily reliable (users may travel or give a false location), our aim was to get a broad overview before the main analysis.}. We located 169k users, then grouped them on the basis of broad geographic region (figure \ref{regional_usage}). Most regions follow the overall trend of using far more light tones. The exception is North America, where the distribution of tones is more uniform---though T5 still has relatively little usage, and is the least common tone worldwide. 
This raises the question of why lighter TME+ are used so much more than darker ones in all regions outside North America. These results would seem to contradict the hypothesis that TME+ are used for self-representation, since we might na{\"i}vely expect more dark-skinned users in Africa and Asia, whereas the TME+ being used are mainly light. 
However, many countries (particularly in Africa) have lower levels of internet access compared to highly developed countries (such as those in North America)\footnote{World Bank, World Telecommunication Report, \url{http://data.worldbank.org/indicator/IT.NET.BBND.P2}}, and access to the internet or modern devices can be even more limited for some ethnic groups due to political oppression \cite{Weidmann1151} or poverty. So, it could simply be that in most regions, lighter-skinned internet users predominate.


Another possible hypothesis for North America's more uniform TME+ distribution (although the demographics do not have similar distributions) is that people use TME+ to refer to both others and themselves. We investigate this hypothesis in the next section, looking more closely at TME usage on Twitter by comparing users' actual skin tone to that seen in their TME+.


\section{Skin Tone Modifiers for Self-Identification}


Since our goal is to better understand how TME relate to user identity, we randomly selected 10k users from Dataset A, subject to the condition that they had used at least one TME. Although we cannot perfectly determine the skin color of these users, we estimate it from their profile photos. We assume that photos of a single individual are self-portraits of the user. This might not be true in all cases, since users can use photos of others. We assume this practice is rare, with a negligible effect on our analysis. However, our annotation process was designed with this possibility in mind.

\begin{figure}
\centering
\includegraphics[width=0.8\columnwidth]{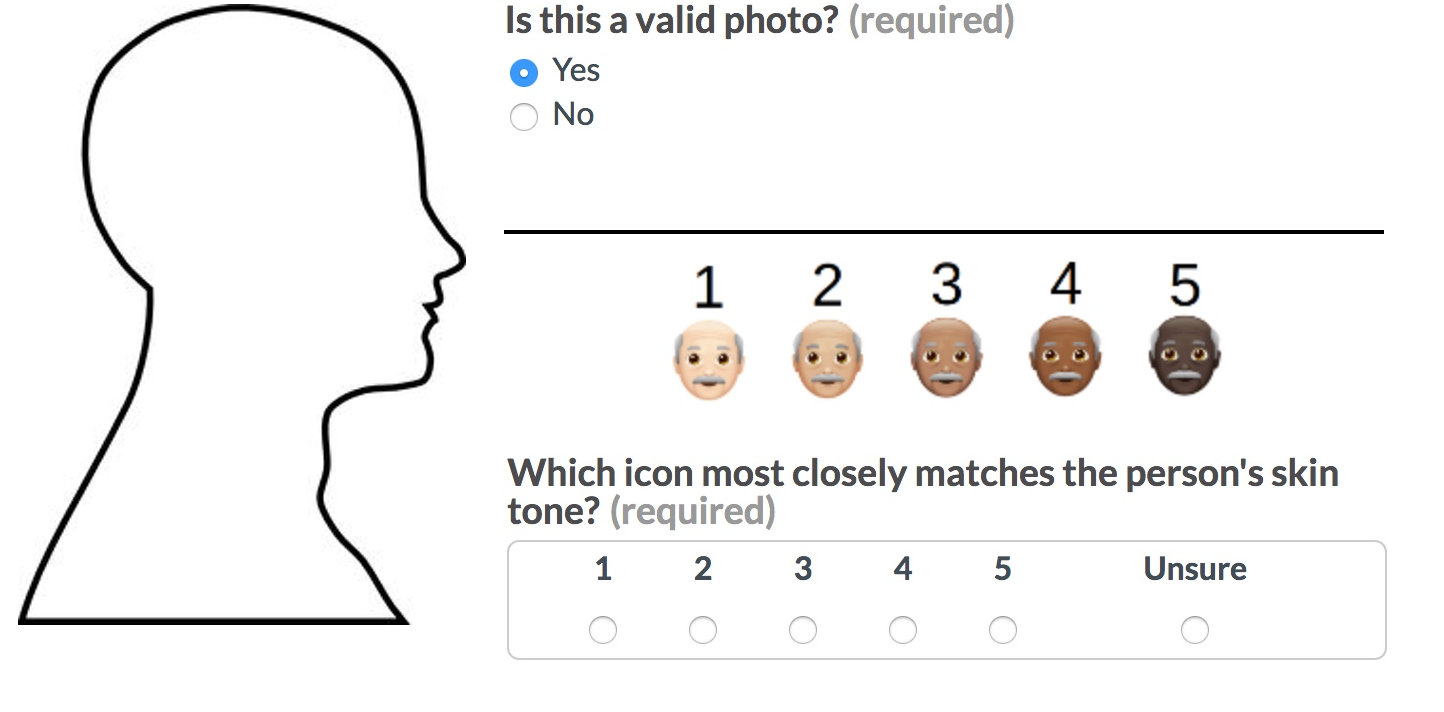}
\caption{Example of Crowdflower annotation task.} 
\label{crowdflowerexample}
\end{figure}

We collected the Twitter profile pictures of the selected 10k users. These pictures were then annotated using Crowdflower\footnote{\url{www.crowdflower.com}}. We instructed annotators to label as ``invalid'' images that might not represent the Twitter user. ``Invalid'' images include: non-human (e.g. animals, nature), celebrities, group photos, gray-scale images. Instructions were given beforehand, with examples of valid/invalid photos provided. If a picture was annotated as ``valid'', annotators compared it to an array of five TME+, as shown in Figure \ref{crowdflowerexample}. We selected the example emoji for its overall lack of distracting features (e.g. little hair covering the face) and consistent appearance (all versions have the same hair/eye color).

Each photo was annotated by at least three annotators. A set of 50 pre-annotated pictures was used to control the quality of the annotators. Only workers achieving over 90\% accuracy on the screening pictures were accepted.

After the annotation process, we retained only those pictures annotated as valid where at least two annotators agreed on the same skin tone. The result was 4,099 annotated pictures. The inter-annotator agreements for these pictures were 98\% for validity, 88\% for user skin tone.

\subsection{TME Usage Analysis}

We collected the historical tweets of the 4,099 annotated users using the Twitter API, to create Dataset B. This contains the tweets of these users from January 2018 back to April 2015---the time when TME were enabled. 5.86m original (non-retweet) tweets were collected in total. 2.46m tweets contained emoji, of which 479k contained TME.

\begin{figure}[t]
\includegraphics[width=\columnwidth]{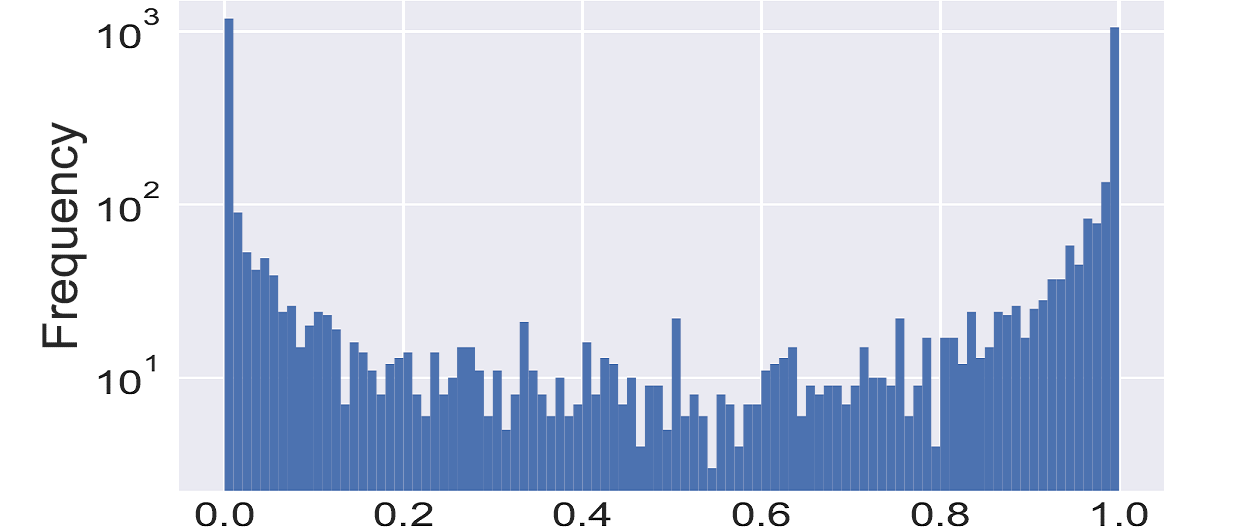}
\caption{Number of users (y-axis, log scale) who use TME+ a given proportion (x-axis) of the time.}
\label{per_user_mean_proportion}
\vspace{10pt}
\includegraphics[width=\columnwidth]{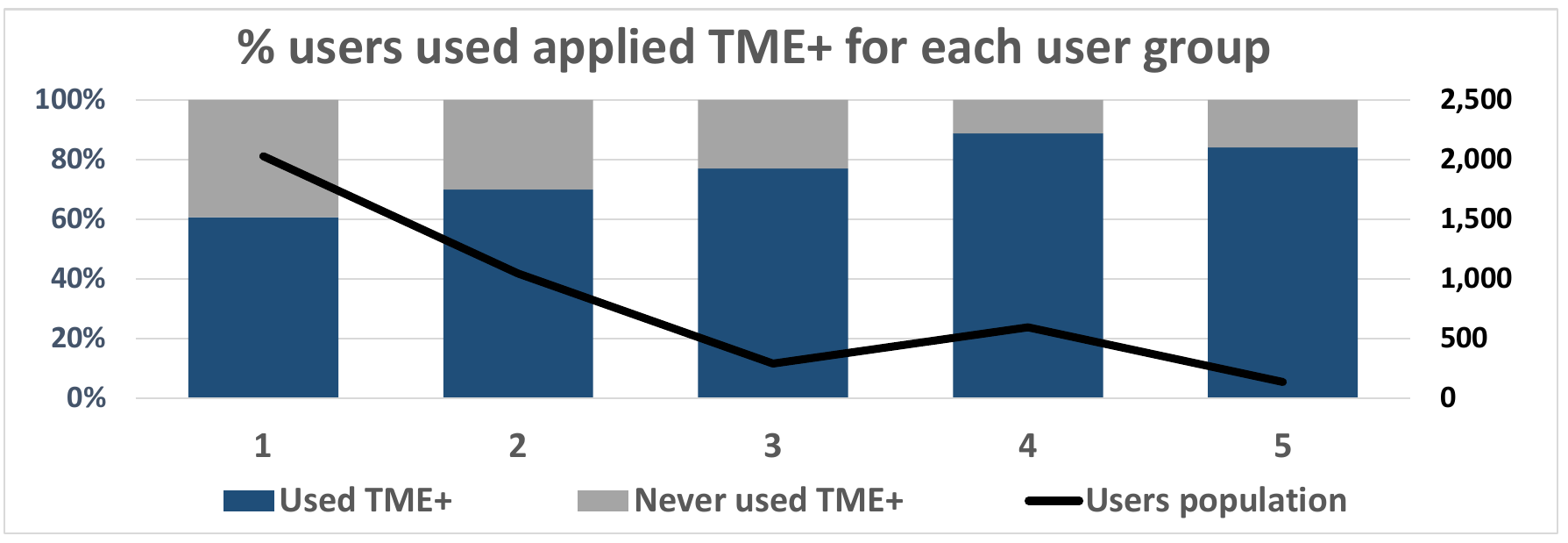}
\caption{Percentage of users in each skin tone group (1=lightest, 5=darkest) who used TME+ at least once. Black line (right-hand axis): number of users per group.}
\label{usersTME}
\vspace{10pt}
\includegraphics[width=\columnwidth]{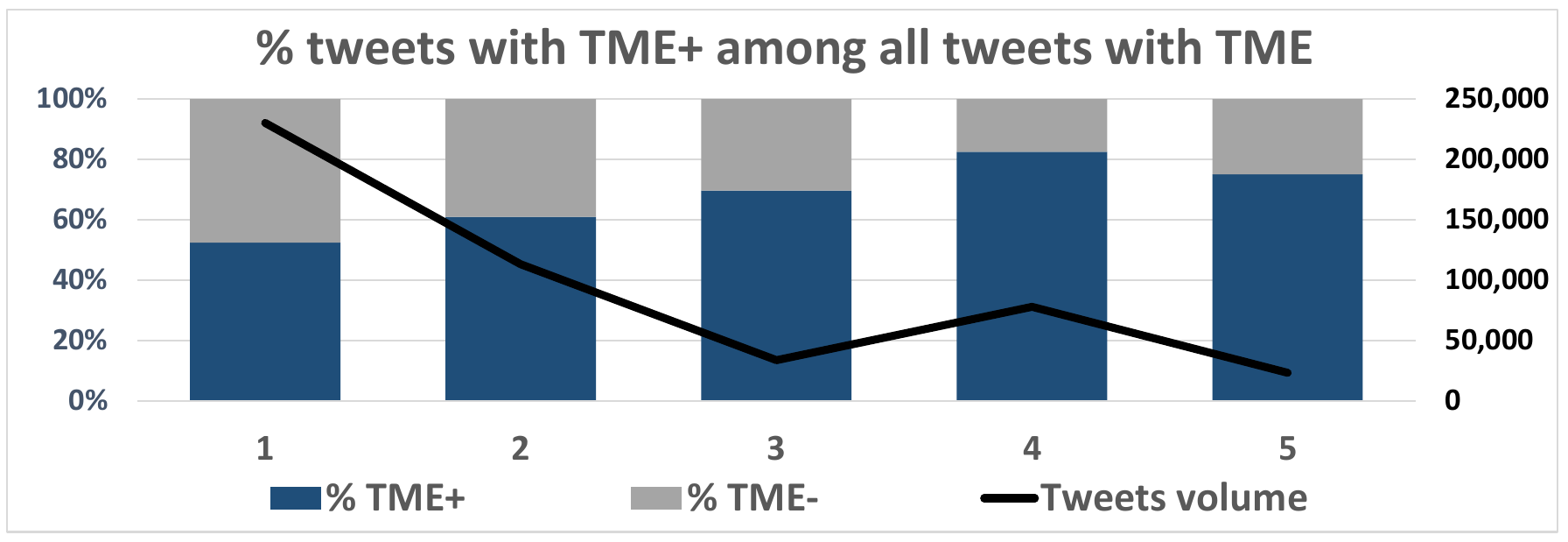}
\caption{Percentage of tweets with TME+ made by users in each skin tone group, out of all tweets with TME. Black line (right-hand axis): number of tweets per group.}
\label{tweetsTME}
\end{figure}

Before looking at the relationship between a user's skin tone and their use of TME, we first examined the overall usage of TME by user. The histogram in Figure \ref{per_user_mean_proportion} shows that, of the Dataset B users, roughly a quarter only ever use TME-, with about the same proportion always using TME+. The remaining users tend towards one end or the other.

Next, we looked at how TME+ usage varies across users according to their annotated skin color. Figure \ref{usersTME} shows the proportion of users in each of the five skin color groups who tweeted at least one TME+. A clear trend is visible, where users with darker skin are more likely to use TME+ than those with lighter skin. In fact, 80+\% of users with skin tone 4 and 5 use TME+ in their tweets, which shows the importance of this feature to them.
 
However, not all TME are TME+. Thus, we examined users who use TME and computed the proportion of those tweets which are TME+. Figure \ref{tweetsTME} shows the percentage of TME+ vs TME- for each user group. Again, the same trend appears: darker skinned users tend to modify the skin tone of TME in their tweets much more than users with lighter skin. For instance, 82\% of TME tweets by users with T4 skin change the skin tone of the emoji compared to only 52\% of the white (T1 skin) users. This emphasizes how users with darker skin are keen to modify the appearance of their emoji, presumably in order to better reflect their own identity.



\subsection{TME+ for Self-Reference vs Referencing Others}

\begin{figure}
\begin{center}
\includegraphics[width=0.6\columnwidth]{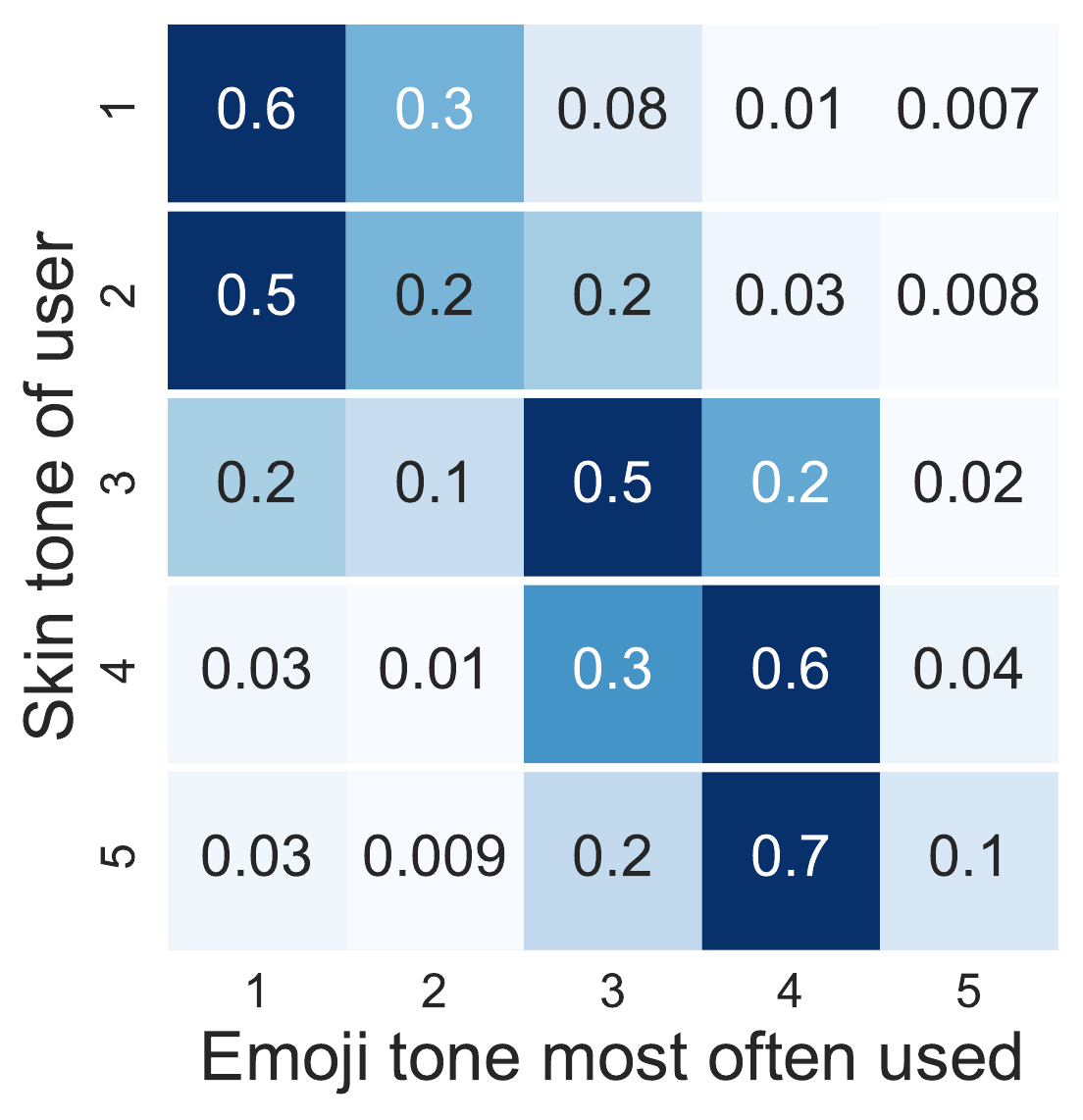}
\caption{Most common emoji skin tone used vs perceived user skin tone, normalized by row sum.}
\label{confusion_matrix}
\end{center}
\end{figure}

Finally, we looked at whether users' choice of modifiers reflects their own skin color or others. For each tone user, we determined the TME+ tone used most often by that user. Figure \ref{confusion_matrix} shows a heatmap of the distribution of the most common skin tones used by each user group in Dataset B.
Overall, our results suggest that by far the most common scenario is for users to choose skin tone modifiers that roughly match their own complexion, accounting for the variability in rendering of these tones across different platforms. There seems to be some tendency for users to choose modifiers towards the lighter end of the available spectrum; for example, users with skin tone 2 mostly select skin tone 1 in their emoji, and users of skin tone 5 mostly select skin tone 4 in their emoji.

\subsection{The (Non)-Prevalence of Digital Blackface}

\begin{figure}
\includegraphics[width=\columnwidth]{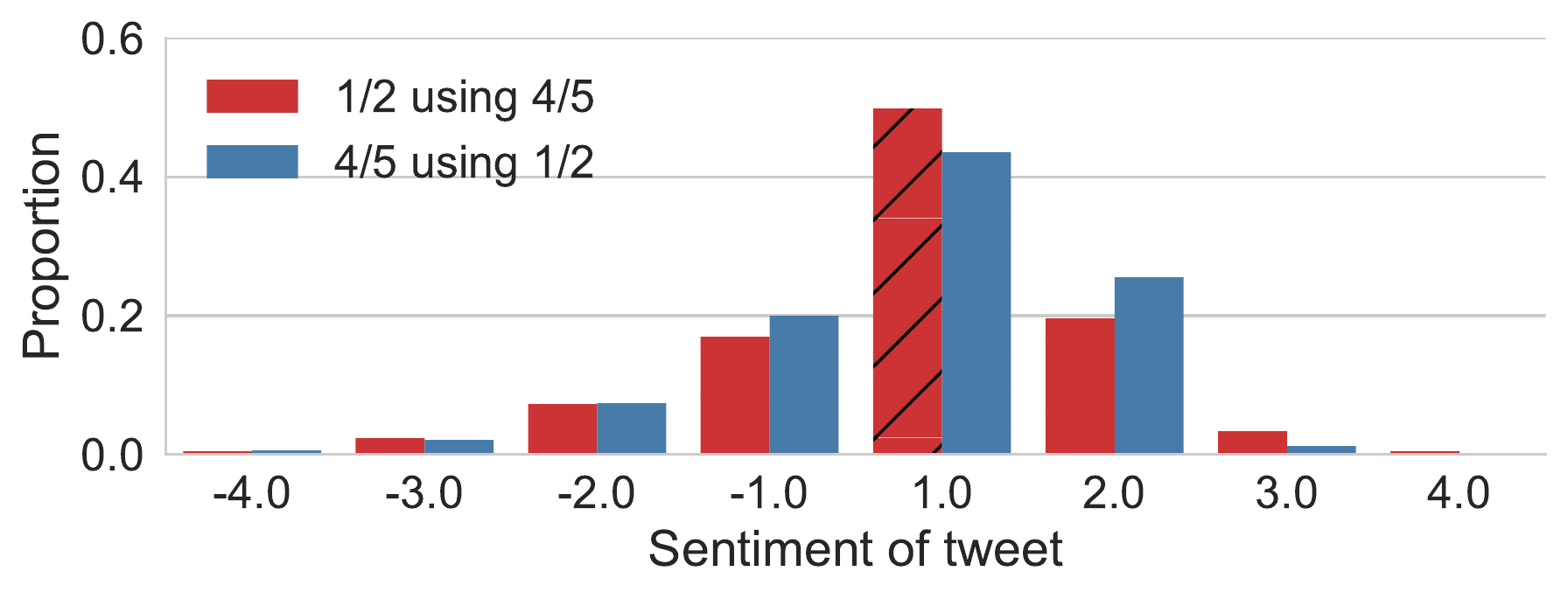}
\caption{Distribution of non-neutral sentiment for English tweets containing lighter/darker TME+ but written by users having darker/lighter skin.}
\label{sentiment_across_groups}
\end{figure}

The previous analysis shows that for nearly all users, their {\em most common} choice of skin tone matches their own skin. However, it still may be the case that users occasionally choose emoji of a different tone. We looked at how often this occurs, and whether there is any noticeable negative sentiment associated with this behavior.
For this analysis, we aggregated Dataset B users into two groups based on skin tone: tones 4/5 (dark) and tones 1/2 (light). First, we analysed the tweets of each group that use TME+ with the skin tone of the other group. These comprise 1229 tweets for the dark-skinned users and 1,729 tweets for the light-skinned users---a very small proportion of each group's tweets: 0.14\% (light group) and 0.24\% (dark group). Next, we examined the opposite-tone tweets in English and computed their sentiment using Sentistrength \cite{thelwall2012sentiment},
which is
designed specifically for analyzing short online messages such as Twitter posts.
The majority of these tweets were neutral (almost 50\% of the tweets of both groups). The distribution of the non-neutral sentiment of tweets for both groups are shown in Figure~\ref{sentiment_across_groups}, where 4 is the most positive and -4 is the most negative\footnote{Sentiment was rated on a -5:+5 scale but no tweets at the extremes of this range were found in this subset of the data.}. As shown, the distributions for both groups are almost the same, and positive tweets outnumber negative ones. Inspection of tweets with negative sentiment revealed tweets on generally negative topics, rather than anything specifically about race. Overall, we find no evidence to justify fears of widespread ``digital blackface'' or its black-against-white counterpart.

\section{Conclusion}
In this paper we presented the first quantitative study on the usage of tone-modifiable emoji (TME) on Twitter. We showed that although lighter-toned emoji are more common overall, different populations of users (based on the skin color of their profile picture) show markedly different characteristics in how they use TME and to what extent. The overall picture is one where users take advantage of emoji skin tone modifiers to represent an important aspect of their identity, and do so differentially depending on their own skin tone: compared to light-skinned users (the majority on Twitter), a higher proportion of dark-skinned users use skin tone modifiers, and they use them more frequently. Here we grouped all geographic regions together in looking at user skin tone, but an interesting question for future work will be to examine the relationship between a user's majority/minority status within their real-life community (i.e., their geographic region) and their use of skin tone modifiers.

\section{Acknowledgements}

This work was supported in part by the EPSRC Centre for Doctoral Training in Data Science, funded by the UK Engineering and Physical Sciences Research Council (grant EP/L016427/1) and the University of Edinburgh.






\bibliography{bibfile.bib}

\begin{thebibliography}{}

\bibitem[\protect\citeauthoryear{Ai \bgroup et al\mbox.\egroup
  }{2017}]{ai2017untangling}
Ai, W.; Lu, X.; Liu, X.; Wang, N.; Huang, G.; and Mei, Q.
\newblock 2017.
\newblock {\em Untangling emoji popularity through semantic embeddings.}
\newblock AAAI press.

\bibitem[\protect\citeauthoryear{Davis and Edberg}{2014}]{tsr51}
Davis, M., and Edberg, P.
\newblock 2014.
\newblock Proposed {D}raft {U}nicode {T}echnical {R}eport: {U}nicode {E}moji.
\newblock Technical Report~51, Unicode Consortium.

\bibitem[\protect\citeauthoryear{Dickey}{2017}]{techcrunchblackface}
Dickey, M.
\newblock 2017.
\newblock Thoughts on white people using dark-skinned emoji.
\newblock {\em TechCrunch}.

\bibitem[\protect\citeauthoryear{Fitzpatrick}{1988}]{fitzpatrick}
Fitzpatrick, T.
\newblock 1988.
\newblock The validity and practicality of sun-reactive skin types {I} through
  {VI}.
\newblock {\em Archives of Dermatology} 124(6):869--871.

\bibitem[\protect\citeauthoryear{Hu \bgroup et al\mbox.\egroup
  }{2017}]{intent2017}
Hu, T.; Guo, H.; Sun, H.; Nguyen, T.~T.; and Luo, J.
\newblock 2017.
\newblock Spice up your chat: {T}he intentions and sentiment effects of using
  emoji.
\newblock {\em CoRR} abs/1703.02860.

\bibitem[\protect\citeauthoryear{Marengo, Giannotta, and
  Settanni}{2017}]{MARENGO201774}
Marengo, D.; Giannotta, F.; and Settanni, M.
\newblock 2017.
\newblock Assessing personality using emoji: {A}n exploratory study.
\newblock {\em Personality and Individual Differences} 112:74 -- 78.

\bibitem[\protect\citeauthoryear{Miller \bgroup et al\mbox.\egroup
  }{2017}]{miller2017}
Miller, H.; Kluver, D.; Thebault-Spieker, J.; Terveen, L.; and Hecht, B.
\newblock 2017.
\newblock {\em Understanding emoji ambiguity in context: {T}he role of text in
  emoji-related miscommunication}.
\newblock AAAI Press.
\newblock  152--161.

\bibitem[\protect\citeauthoryear{Princewill}{2017}]{bbcblackface}
Princewill, V.
\newblock 2017.
\newblock Is it {OK} to use black emojis and gifs?
\newblock {\em BBC News}.

\bibitem[\protect\citeauthoryear{Thelwall, Buckley, and
  Paltoglou}{2012}]{thelwall2012sentiment}
Thelwall, M.; Buckley, K.; and Paltoglou, G.
\newblock 2012.
\newblock Sentiment strength detection for the social web.
\newblock {\em Journal of the Association for Information Science and
  Technology} 63(1):163--173.

\bibitem[\protect\citeauthoryear{Weidmann \bgroup et al\mbox.\egroup
  }{2016}]{Weidmann1151}
Weidmann, N.~B.; Benitez-Baleato, S.; Hunziker, P.; Glatz, E.; and
  Dimitropoulos, X.
\newblock 2016.
\newblock Digital discrimination: {P}olitical bias in internet service
  provision across ethnic groups.
\newblock {\em Science} 353(6304):1151--1155.

\bibitem[\protect\citeauthoryear{Wijeratne \bgroup et al\mbox.\egroup
  }{2017}]{emojinet2017}
Wijeratne, S.; Balasuriya, L.; Sheth, A.~P.; and Doran, D.
\newblock 2017.
\newblock Emoji{N}et: {A}n open service and {API} for emoji sense discovery.
\newblock In {\em ICWSM},  437--447.

\bibitem[\protect\citeauthoryear{Zimmerman}{2015}]{guardiandiverse}
Zimmerman, J.
\newblock 2015.
\newblock Racially diverse emoji are a nice idea. {B}ut will anyone use them?
\newblock {\em The Guardian}.

\end{thebibliography}
\bibliographystyle{aaai}
\end{document}